# Single-ion versus two-ion anisotropy in magnetic compounds: A neutron scattering study

A. Furrer[1], F. Juranyi[1], K. W. Krämer[2], M. Schneider[1], Th. Strässle[1]

[1] Laboratory for Neutron Scattering, Paul Scherrer Institut, CH-5232 Villigen PSI, Switzerland

[2] Department of Chemistry and Biochemistry, University of Bern, CH-3012 Bern, Switzerland

**Abstract:**

Anisotropy effects can significantly control or modify the ground-state properties of magnetic systems. Yet the origin and the relative importance of the possible anisotropy terms is difficult to assess experimentally and often ambiguous. Here we propose a technique which allows a very direct distinction between single-ion and two-ion anisotropy effects. The method is based on high-resolution neutron spectroscopic investigations of magnetic cluster excitations. This is exemplified for manganese dimers and tetramers in the mixed compounds $CsMn_xMg_{1-x}Br_3$ (0.05≤x≤0.40). Our experiments provide evidence for a pronounced anisotropy of the order of 3% of the dominant bilinear exchange interaction, and the anisotropy is dominated by the single-ion term. The detailed characterization of magnetic cluster excitations offers a convenient way to unravel anisotropy effects in any magnetic material.

# I. Introduction

The properties of magnetic systems are commonly interpreted in terms of the Heisenberg Hamiltonian $H=-2\Sigma_{i>j}J_{ij}\mathbf{s_i}\cdot\mathbf{s_j}$ where $\mathbf{s_i}$ is a spin operator and $J_{ij}$ an isotropic bilinear exchange parameter which couples the magnetic ions at sites i and j. The Heisenberg model, however, represents a simplification of the true situation, since anisotropy is always present in real materials at some energy scale, and it can significantly modify the magnetic ground-state properties [1]. Often on account of experimental findings additional terms have to be added to the Heisenberg Hamiltonian such as single-ion anisotropy [2], symmetric and antisymmetric exchange anisotropies [3], and higher-order exchange interactions [4]. The parameters of the spin Hamiltonian are usually derived by combining theoretical relations for various magnetic properties, notably the spin-wave dispersion, with experimental data on these properties. This strategy is not always successful, since for three-dimensional systems exact solutions cannot be obtained. In particular, the commonly used formula of the spin-wave dispersion is an approximation based on the linearization of the equation of motion for the spin operators. Moreover, the spin-wave dispersion does not allow the separate determination of all the individual coupling terms, so that, *e.g.*, the relative size of the bilinear exchange terms with respect to the higher-order coupling terms cannot be assessed [4]. Similarly, the distinction between single-ion and two-ion anisotropy terms is often ambiguous. These difficulties can be overcome by studying diluted systems, in which small clusters of exchange-coupled magnetic ions occur in isolation, so that the spin Hamiltonian can be solved exactly allowing a rigorous comparison between theory and experiment.

      The present work addresses primarily the question how relatively weak single-ion and two-ion anisotropies can be distinguished from each other by studying magnetic cluster excitations. As model systems we used mixed

compounds of composition $CsMn_xMg_{1-x}Br_3$ (0.05≤x≤0.40) for various reasons. Both $CsMnBr_3$ and $CsMgBr_3$ crystallize in the hexagonal space group $P6_3/mmc$, and their unit cell parameters are almost identical: a=b=7.609(15) Å, c=6.52(5) Å for $CsMnBr_3$ [5] and a=b=7.610(2) Å, c=6.502(2) Å for $CsMgBr_3$ [6]. The structure consists of chains of face-sharing $MBr_6$ (M=$Mn^{2+}$, $Mg^{2+}$) octahedra parallel to the c axis. Spin-wave experiments on $CsMnBr_3$ gave evidence for a pronounced one-dimensional magnetic behavior with the intrachain exchange interaction exceeding the interchain exchange interaction by three orders of magnitude [7,8]. All the $Mn^{2+}$ clusters in the mixed compounds $CsMn_xMg_{1-x}Br_3$ are thus linear chain fragments with composition $Mn_nBr_{3(n+1)}$ (n=1,2,3,…) oriented parallel to the c axis. Inelastic neutron scattering experiments on $Mn^{2+}$ dimers (n=2) and trimers (n=3) showed that the biquadratic exchange interaction distinctly contributes to the spin coupling of the $Mn^{2+}$ ions [4,9], but there was no evidence for the presence of an anisotropy term which resulted from the analysis of the spin-wave experiments. With increased instrumental resolution used in the present work, however, we were able to detect anisotropy-induced splittings of magnetic cluster excitations, and the combined analysis of some dimer and tetramer transitions resulted in an unambiguous assessment of the nature of the underlying anisotropies.

The present work is organized as follows. The experimental procedure is described in Sec. II, followed in Sec. III by a summary of the spin Hamiltonians and neutron cross-sections for spin dimers and tetramers. In addition, numerical values are tabulated for the singlet-triplet transitions as well as for the anisotropy-induced triplet splittings in antiferromagnetically coupled dimers and tetramers of transition metal ions with spin quantum numbers $1/2 \leq s_i \leq 5/2$. The experimental results and their analyses are presented in Sec. IV. Finally, some conclusions are given in Sec. V. A statistical model addressing the linewidth of dimer excitations is described in the Appendix.

## II. EXPERIMENT

Polycrystalline samples of $CsMn_xMg_{1-x}Br_3$ (x=0.05, 0.10, 0.14, 0.28, 0.40) were synthesized according to standard procedures [4]. The inelastic neutron scattering experiments were carried out with use of the high-resolution time-of-flight spectrometer FOCUS at the spallation neutron source SINQ at PSI Villigen. The measurements were performed with incoming neutron energies of 5.11 and 2.91 meV in the time-focusing mode, which minimizes the instrumental energy resolution at the energy transfer of interest. The scattered neutrons were detected by an array of $^3$He counters covering a large range of scattering angles $10°≤\Phi≤130°$. The samples were enclosed in Al cylinders (12 mm diameter, 45 mm height) and placed into a He cryostat to achieve temperatures $1.5≤T≤50$ K. Additional experiments were performed for the empty container as well as for vanadium to allow the correction of the raw data with respect to background, detector efficiency, absorption, and detailed balance according to standard procedures.

## III. THEORETICAL BACKGROUND

### A. Dimer Excitations

We base the analysis of the dimer transitions on the spin Hamiltonian

$$H = -2J\mathbf{s_1}\cdot\mathbf{s_2} - K(\mathbf{s_1}\cdot\mathbf{s_2})^2 - 2J^z s_1^z s_2^z - D\left[(s_1^z)^2 + (s_2^z)^2\right] \quad (1)$$

where $\mathbf{s_i}$ denotes the spin operator of the magnetic ions, J and K the bilinear and the biquadratic exchange interaction, respectively, and $J^z$ and D the two-ion and the single-ion anisotropy parameter, respectively. The particular choice of the anisotropy terms is dictated by the axial symmetry of the $CsMn_xMg_{1-x}Br_3$ compounds. H commutes with the total spin $\mathbf{S}=\mathbf{s_1}+\mathbf{s_2}$, thus S is a good quantum number to describe the spin states as |S,M> with -S≤M≤S. For $J^z$=D=0 and identical magnetic ions ($s_1=s_2$) the eigenvalues of Eq. (1) are degenerate with respect to the quantum number M:

$$E(S) = -J\eta - \frac{1}{4}K\eta^2 \; , \; \eta = S(S+1) - 2s_i(s_i+1) \; , \; 0 \leq S \leq 2s_i \qquad (2)$$

For antiferromagnetic exchange (J<0) the ground state is a singlet (S=0), separated from the excited triplet (S=1), quintet (S=2), *etc.* states according to the well-known Landé interval rule which for K=0 is given by

$$E(S) - E(S-1) = -2JS \qquad (3)$$

Non-zero anisotropy terms ($J^z$≠0 and/or D≠0) have the effect of splitting the spin states |S> into the states |S,±M>. For instance, for D>0 and $J^z$>0 the excited triplet (S=1) is split into a lower-lying doublet |1,±1> and a higher lying singlet |1,0>, whereas for D<0 and $J^z$<0 the energetic ordering of the two sublevels is reversed. This is exemplified for dimers with $s_i$=5/2 in Fig. 1 which also includes the splitting of the excited quintet state (S=2). From the observation of the triplet (S=1) splitting alone the nature of the anisotropy cannot be determined, but the anisotropy induced splitting of the quintet (S=2) is sufficiently detailed to arrive at a distinction between the parameters D and $J^z$.

For spin dimers the neutron cross-section for a transition from the initial state |S> to the final state |S'> is defined by [10]

$$\frac{d^2\sigma}{d\Omega d\omega} = \frac{N}{Z}(\gamma r_0)^2 \frac{k'}{k} F^2(\mathbf{Q}) \exp\{-2W(\mathbf{Q})\} \exp\left\{-\frac{E(S)}{k_B T}\right\} \sum_\alpha \left[1 - \left(\frac{Q_\alpha}{Q}\right)^2\right]$$

$$\times \frac{2}{3}\left[1 - (-1)^{\Delta S} \cos(\mathbf{Q}\cdot\mathbf{R})\right] |T_1|^2 \delta\{\hbar\omega + E(S) - E(S')\}$$ (4)

where N is the total number of spin dimers in the sample, Z the partition function, k and k' the wave numbers of the incoming and scattered neutrons, respectively, $\mathbf{Q}=\mathbf{k}-\mathbf{k'}$ the scattering vector, $F(\mathbf{Q})$ the magnetic form factor, $\exp\{-2W(\mathbf{Q})\}$ the Debye-Waller factor, $\mathbf{R}$ the vector defining the intradimer separation, $T_i = \langle S' \| T_i \| S \rangle$ ($T_1 = T_2$) the reduced transition matrix element defined in Ref. 10, and $\hbar\omega$ the energy transfer. The remaining symbols have their usual meaning. The transition matrix element carries essential information to derive the selection rules for spin dimers:

$$\Delta S = S - S' = 0, \pm 1 \ ; \ \Delta M = M - M' = 0, \pm 1 \ .$$ (5)

Eq. (4) is valid as long as the states $|S,M\rangle$ are degenerate with respect to M. For polycrystalline material Eq. (4) has to be averaged in $\mathbf{Q}$ space. By separating the $\Delta M=0$ and $\Delta M=\pm 1$ transitions we obtain [11]

$$\left.\frac{d^2\sigma}{d\Omega d\omega}\right|_{\Delta M=0} \propto F^2(Q)\left\{\frac{2}{3} + (-1)^{\Delta S}\left[\frac{2\sin(QR)}{(QR)^3} - \frac{2\cos(QR)}{(QR)^2}\right]\right\}|T_1^{\Delta M=0}|^2$$

(6)

$$\left.\frac{d^2\sigma}{d\Omega d\omega}\right|_{\Delta M=\pm 1} \propto F^2(Q)\left\{\frac{2}{3} - (-1)^{\Delta S}\left[\frac{\sin(QR)}{(QR)^3} - \frac{2\cos(QR)}{(QR)^2} - \frac{\sin(QR)}{QR}\right]\right\}|T_1^{\Delta M=\pm 1}|^2$$

For small anisotropy parameters $|D|$, $|J^z| \ll |J|$ each matrix element $|T_1^{\Delta M}|^2$ corresponds to one third of the matrix element $|T_1|^2$ in Eq. (4).

## B. Tetramer Excitations

We base the analysis of transitions associated with linear tetramers on the spin Hamiltonian

$$H = -2J(\mathbf{s}_1 \cdot \mathbf{s}_2 + \mathbf{s}_2 \cdot \mathbf{s}_3 + \mathbf{s}_3 \cdot \mathbf{s}_4) - K\left[(\mathbf{s}_1 \cdot \mathbf{s}_2)^2 + (\mathbf{s}_2 \cdot \mathbf{s}_3)^2 + (\mathbf{s}_3 \cdot \mathbf{s}_4)^2\right]$$
$$- 2J^z(s_1^z s_2^z + s_2^z s_3^z + s_3^z s_4^z) - D\left[(s_1^z)^2 + (s_2^z)^2 + (s_3^z)^2 + (s_4^z)^2\right] \quad (7)$$

To solve Eq. (7) the total spin S defined by $\mathbf{S}=\mathbf{s}_1+\mathbf{s}_2+\mathbf{s}_3+\mathbf{s}_4$ is still a good quantum number, but for a complete characterization of the tetramer states additional spin quantum numbers are needed, *e.g.*, $\mathbf{S}_{12}=\mathbf{s}_1+\mathbf{s}_2$ and $\mathbf{S}_{34}=\mathbf{s}_3+\mathbf{s}_4$ with $0 \leq S_{12} \leq 2s_i$ and $0 \leq S_{34} \leq 2s_i$, respectively. The total spin is then defined by $|S_{12}-S_{34}| \leq S \leq (S_{12}+S_{34})$, and the basis states are given by the wavefunction $|S_{12},S_{34},S\rangle$. There is no spin coupling scheme which results in a diagonal energy matrix, so that the eigenvalues of Eq. (7) have to be calculated by conventional spin operator techniques [12]. For antiferromagnetic exchange J<0 the ground state is a singlet (S=0) defined by the wavefunction $|\Psi_0\rangle = \Sigma_i \alpha_i |S_{12}(i),S_{34}(i),0\rangle$ with $\Sigma_i(\alpha_i)^2=1$. It can be shown that for any spin quantum number $s_i$ the first excited state is always a triplet (S=1) defined by the wavefunction $|\Psi_1\rangle = \Sigma_i \beta_i |S_{12}(i),S_{34}(i),1\rangle$ with $\Sigma_i(\beta_i)^2=1$, which for non-zero anisotropy terms ($J^z \neq 0$ and/or $D \neq 0$) is split into a doublet (M=±1) and a singlet (M=0) as for the dimer case discussed in Sec. III.A.

The cross-section for the tetramer transition $|S_{12},S_{34},S\rangle \rightarrow |S'_{12},S'_{34},S'\rangle$ takes the form [13]

$$\frac{d^2\sigma}{d\Omega d\omega} = \frac{N}{Z}(\gamma r_0)^2 F^2(\mathbf{Q})\exp\{-2W(\mathbf{Q})\}$$

$$\times \exp\left\{-\frac{E(S_{12},S_{34},S)}{k_B T}\right\}\sum_\alpha\left[1-\left(\frac{Q_\alpha}{Q}\right)^2\right]$$

$$\times \frac{2}{3}\left\{\delta(S_{34},S'_{34})\left[1-(-1)^{S_{12}-S'_{12}}\cos(\mathbf{Q}\cdot\mathbf{R_{12}})\right]|T_1|^2\right.$$

$$+ \delta(S_{12},S'_{12})\left[1-(-1)^{S_{34}-S'_{34}}\cos(\mathbf{Q}\cdot\mathbf{R_{34}})\right]|T_3|^2 \quad (8)$$

$$+ \delta(S_{12},S'_{12})\delta(S_{34},S'_{34})\left[\cos(\mathbf{Q}\cdot\mathbf{R_{13}})\right.$$

$$\left.+\cos(\mathbf{Q}\cdot\mathbf{R_{14}})+\cos(\mathbf{Q}\cdot\mathbf{R_{23}})+\cos(\mathbf{Q}\cdot\mathbf{R_{24}})\right]T_1 T_3\bigg\}$$

$$\times \delta\{\hbar\omega + E(S_{12},S_{34},S) - E(S'_{12},S'_{34},S')\}$$

where N is the total number of tetramers, $\mathbf{R}_{ij}$ the distance vector between the magnetic ions at sites i and j, $T_i = \langle S'_{12},S'_{34},S\|T_i\|S_{12},S_{34},S\rangle$ the reduced transition matrix element ($T_1=T_2$, $T_3=T_4$), and the remaining symbols are as in Eq. (4). From the reduced matrix elements the following selection rules are obtained:

$$\Delta S_{12} = 0, \pm 1 \; ; \; \Delta S_{34} = 0, \pm 1 \; ; \; \Delta S = 0, \pm 1 \; ; \; \Delta M = 0, \pm 1 \quad (9)$$

**C. Properties of the dimer and linear tetramer excitations**

Both magnetic dimer and linear tetramer systems have a singlet (S=0) ground state for antiferromagnetic exchange coupling (J<0) in Eqs (1) and (7), respectively, and the first excited state is always a triplet (S=1). For dimers the separation between the singlet and the triplet is $\Delta_d=-2J$ independent of the spin quantum number $s_i$ of the individual magnetic ions. The singlet-triplet splitting $\Delta_t$ of linear tetramers is always smaller than $\Delta_d$, but its size depends on $s_i$ as listed in Table I. As mentioned in Sections III.A and III.B, the doublet |1,±1⟩ lies below the singlet |1,0⟩ for D>0 and $J^z$>0. For both dimers and tetramers, the

triplet splittings δ are additive, *i.e.*, anisotropy parameters D and $J^z$ with different signs can largely compensate each other.

## IV. RESULTS AND DATA ANALYSIS

Energy spectra of neutrons scattered from $CsMn_xMg_{1-x}Br_3$ at T=1.5 K are shown for different Mn concentrations x in Fig. 2. The spectrometer parameters were chosen to achieve an optimal instrumental resolution of 55 μeV at an energy transfer of $\Delta_d \approx 1.8$ meV where the dimer |0>→|1> excitation is expected to occur. There are two well defined lines at 1.80 and 1.93 meV, which we attribute to the dimer excitations |0,0>→|1,±1> and |0,0>→|1,0>, respectively. This identification is supported by the Q-dependence of the intensities displayed in Fig. 3, which compares the observed intensities with those calculated from the cross section (6). The splitting of the excited dimer triplet due to anisotropy effects turns out to be independent of x (indicated by the dashed-dotted lines in Fig. 2) and amounts to $\delta_d$=0.135(3) meV, which can be rationalized either with a single-ion anisotropy parameter D=0.0211(5) meV or with a two-ion anisotropic exchange parameter $J^z$=0.0183(4) meV or with a linear combination $yD+(1-y)J^z$ where y can take any value. These parameters were obtained from a least-squares fitting procedure in which the bilinear exchange parameter J was also allowed to vary, whereas the biquadratic exchange parameter was held constant at the temperature-independent value K=8.6(2) μeV [9]. We found J=-0.852(3) meV (at T=1.5 K), which slightly exceeds the values J=-0.838(5) meV (for 2≤T≤70 K) [4] and J=-0.823(1) meV (at T=50 K) [9] determined from the dimer excitations taken at higher temperatures and in the absence of anisotropy terms.

  The linewidth of the excitations considerably depends on the concentration x as demonstrated in Fig. 4, which shows the intrinsic linewidths corrected for the instrumental resolution. With increasing x the linewidth is enhanced due to inhomogeneities along the mixed $Mn_xMg_{1-x}$ chains resulting from the different ionic radii of the $Mn^{2+}$ and $Mg^{2+}$ ions. The x-dependence of the linewidth nicely follows a $\sigma^2$ law where $\sigma$ is the variance of the probabilities $p_m(x)$ (m=0,1,2,...) for having m $Mn^{2+}$ ions on both sides of the central $Mn^{2+}$ pair in a mixed $Mn_xMg_{1-x}$ chain as outlined in detail in the Appendix.

The instrumental setting used to collect the data of Fig. 2 also provided data at lower energy transfers around $\Delta_t \approx 0.9$ meV where the tetramer transition |0>→|1> is expected to occur, with an instrumental resolution of about 110 µeV. For Mn concentrations x=0.05, 0.10 and 0.14, the probabilty for $Mn^{2+}$ tetramer formation is less than 0.23 %, so that no relevant signal could be detected. For x=0.28 and 0.40, however, the probability for tetramer formation is drastically enhanced to 1.58 and 3.84 %, respectively. Fig. 5 shows energy spectra taken for x=0.28 and 0.40 with subtraction of the x=0.14 data, which has the advantage that uncertainties about the background are automatically eliminated. The observed overall intensity ratio I(x=0.40)/I(x=0.28)=2.8(3) is in agreement with the corresponding probabilities for tetramer formation whose ratio is 2.43. There are two bands at 0.79 and 0.98 meV, which we attribute to the tetramer excitations |0,0>→|1,±1> and |0,0>→|1,0>, respectively. This identification is supported by comparing the intensities with the cross section (8). The tetramer |0>→|1> transitions are governed by the selection rules $\Delta S_{12}=0$, $\Delta S_{34}=\pm 1$ or $\Delta S_{12}=\pm 1$, $\Delta S_{34}=0$, but in no case transitions with both $\Delta S_{12}=\pm 1$ and $\Delta S_{34}=\pm 1$ occur, so that the cross section (8) reduces to the form displayed for dimers in Fig. 3. The calculated intensity ratio I($\Delta M=\pm 1$)/I($\Delta M=0$)=3.0 is in good agreement with the observed ratio of 3.2(4). We therefore conclude that the splitting of the excited tetramer triplet due to anisotropy effects is $\delta_t$=0.196(9) meV, which can be rationalized either with a single-ion anisotropy parameter D=0.0224(10) meV or with a two-ion anisotropic exchange parameter $J^z$=0.0116(5) meV or with a linear combination $yD+(1-y)J^z$. The above anisotropy parameters were obtained by keeping J=-0.852 meV and K=8.6 µeV fixed at the corresponding dimer values.

The linewidth of the tetramer transitions is increasing with the concentration x, giving intrinsic linewidths of 60(20) and 90(10) µeV for x=0.28 and 0.40, respectively. The intrinsic linewidths of the tetramer transitions are found to be considerably smaller than those of the dimer transitions displayed in Fig. 4. Obviously the spin tetramers are more stable against structural inhomogeneities along the $Mn_xMg_{1-x}$ chain.

The joint analysis of the dimer and tetramer data allows us now to determine the nature of the observed anisotropy. Fig. 6 shows a plot of the parameter values D and $J^z$ which are compatible with the observed triplet splittings $\delta_d$ and $\delta_t$ on the basis of the data for $s_i$=5/2 listed in Table I. The two lines cross at the parameter values

D=0.0193(23) meV, $J^z$=0.0015(19) meV,

*i.e.*, the single-ion anisotropy is to a large extent the origin of the observed triplet splittings.

The observation of the dimer quintet (S=2) splitting offers an alternative way to determine the anisotropy parameters D and $J^z$ separately. Experimentally, the quintet state can be accessed by transitions out of the excited triplet (S=1) state. According to Eq. (3) the transition energy is around -4J≈3.6 meV. Fig. 7 shows an energy spectrum of neutrons scattered from $CsMn_{0.14}Mg_{0.86}Br_3$ at T=25 K corresponding to the excited dimer |1>→|2> transition, which has a maximum around 3.6 meV and a slight asymmetry on the high-energy side. The optimal instrumental resolution of about 120 µeV, which can be achieved on the spectrometer FOCUS, is not sufficient to resolve the five allowed triplet-quintet transitions displayed in Fig. 1, but the skewness of the observed peak offers a convenient means to distinguish between D and $J^z$. The skewness s is defined by the third moment of an energy distribution:

$$s = \sum_i I_i \left( \frac{E_i - <E>}{\sigma} \right)^3 \qquad (10)$$

where $I_i$ is the intensity at the energy transfer $E_i$, $<E>$ the mean energy, and $\sigma$ the variance. Zero skewness corresponds to a symmetric peak, whereas negative and positive values of the skewness indicate asymmetries on the low-energy and high-energy side, respectively. The data displayed in Fig. 7 result in a skewness s=0.155(19) with $<E>$=3.635(2) meV and $\sigma$=0.084(2) meV. In Fig. 8 we calculate the skewness of the dimer $|1> \rightarrow |2>$ transition for the same $(D,J^z)$ pairs as in Fig. 6 which are compatible with the observed splitting $\delta_d$=0.135 meV of the dimer triplet state. For s=0.155(19) we derive the following anisotropy parameters from Fig. 8:

D=0.0173(24) meV, $J^z$=0.0030(20) meV,

which are in good agreement with those derived independently from the joint analysis of the dimer and tetramer triplet splittings $\delta_d$ and $\delta_t$, respectively. The present neutron spectroscopic data are thus best described by the parameters

J=-0.852(3) meV, K=0.0086(2) meV, D=0.0183(16) meV, $J^z$=0.0022(14) meV.

## V. CONCLUSIONS

The anisotropy of the magnetic interactions in $CsMn_xMg_{1-x}Br_3$ was unravelled in a comprehensive neutron spectroscopic study of low-lying $Mn^{2+}$ dimer and tetramer excitations. The observed anisotropy was shown to be predominantly of a single-ion origin due to the axial ligand field with an anisotropy parameter D=0.0183(16) meV, which is considerably larger than the value D=0.012(1) meV determined from the analysis of the spin-wave dispersion [8]. We consider our anisotropy parameter D to be more reliable, since it results directly from the observed splitting energies, independent of the other parameters of the spin Hamiltonian, whereas in the spin-wave formalism the parameters D and J enter as products [8]. Likewise, electron spin resonance experiments which are considered to be a powerful method to determine anisotropy effects, can only detect splittings of individual spin multiplets ($\Delta S=0$) [1], whereas neutron spectroscopy offers in addition the observation of splittings produced by the magnetic interactions ($\Delta S=\pm 1$).

In real magnets, the dipole-dipole interaction is always present in addition to the exchange interaction. The corresponding Hamiltonian for a spin dimer is composed of an isotropic and an anisotropic term:

$$H = \frac{g^2\mu_B^2}{R^3}\left[\mathbf{s_1}\cdot\mathbf{s_2} - 3\frac{(\mathbf{s_1}\cdot\mathbf{R})(\mathbf{s_2}\cdot\mathbf{R})}{R^2}\right] \quad , \tag{11}$$

where g is the Landé splitting factor and $\mu_B$ the Bohr magneton. For $Mn^{2+}$ dimers in $CsMn_xMg_{1-x}Br_3$ with g=2 and R=3.25 Å, the isotropic dipole-dipole coupling parameter amounts to $-(g\mu_B)^2/2R^3$=-0.0063 meV, which is more than two orders of magnitude smaller than the isotropic exchange parameter J=-0.852(3) meV. The exchange coupling J is sufficiently strong to keep the spins $\mathbf{s_1}$ and $\mathbf{s_2}$ antiferromagnetically aligned at low temperatures T≤|J|/$k_B$, but their direction with respect to $\mathbf{R}$//c is free to rotate. Therefore, the second part of Eq. (11) has to be averaged in space, giving rise to an anisotropic dipole-dipole coupling parameter $3(g\mu_B/\pi)^2/2R^3$=0.0019 meV which is presumably the origin of the anisotropic exchange parameter $J^z$=0.0022(14) meV determined in the present work.

Our study was focussed on the investigation of $Mn^{2+}$ dimers and tetramers. This choice is motivated by the nature of the ground-state which for antiferromagnetically coupled dimers and tetramers is always a singlet (S=0). Antiferromagnetically coupled spin trimers are formed in mixed compounds as well, with a larger probability than tetramers, but their ground-state is never a singlet. In fact, for the present case the ground-state of $Mn^{2+}$ trimers is a sextet (S=5/2) [4], giving rise to a large number of transitions to the higher-lying states which can hardly be separated from each other in neutron spectroscopic experiments.

The cluster method introduced in the present work can easily be adapted to anisotropy terms different from those used in the spin Hamiltonians (1) and (7). Each anisotropy term produces its specific splitting pattern of the spin states |S> which allows a rigorous distinction. For instance, a planar single-ion anisotropy of the form $E[(s_i^x)^2+(s_i^y)^2]$ or an antisymmetric two-ion interaction described by the vector product $\mathbf{s_i} \times \mathbf{s_j}$ [14,15] are often relevant in currently studied materials, notably in quantum spin systems like spin-ladder materials [1], in giant magnetoresistance manganates [16], in cobaltates [17], and in single-molecule magnets [18]. The only requirement to apply the cluster method is the existence of mixed compounds in which the magnetic ions are partially substituted by non-magnetic ions, however, chemistry is rich to provide such materials as demonstarted *e.g.* for the mixed manganese compounds $LaMn_xGa_{1-x}O_3$ ($0 \leq x \leq 1$) [19].

## ACKNOWLEDGMENTS

This work was performed at the Swiss Spallation Neutron Source SINQ, Paul Scherrer Institut (PSI), Villigen, Switzerland.

## APPENDIX: LINEWIDTH OF DIMER EXCITATIONS

The intrinsic linewidth of the dimer excitations results from structural inhomogeneities along the mixed $Mn_xMg_{1-x}$ chain, since the ionic radii of the $Mn^{2+}$ and $Mg^{2+}$ ions are different with $r_{Mn}$=83 pm > $r_{Mg}$=72 pm [20]. In Fig. 9 we consider different configurations along the chain, where m is the number of $Mn^{2+}$ ions replacing the $Mg^{2+}$ ions. The introduction of additional $Mn^{2+}$ ions exerts some internal pressure within the chain, so that the atomic positions have to rearrange. In particular, the Mn-Mn bond distance R of the central $Mn^{2+}$ pair will be gradually shortened with increasing number m of $Mn^{2+}$ ions as compared to the case m=0. The intradimer exchange interaction J was shown to vary with R according to d|J|/dR=-3.6(3) meV/Å [9], thus any change of R results in a corresponding change of J and thereby in a line broadening.

Assuming a statistical distribution of $Mn^{2+}$ ions, the probabilities $p_m(x)$ for having m $Mn^{2+}$ ions on both sides of the central $Mn^{2+}$ pair in a chain of length 2n are given by

$$p_0(x) = (1-x)^{2n}$$

$$p_1(x) = 2\binom{n}{1}x(1-x)^{2n-1}$$

$$p_2(x) = \left[2\binom{n}{2} + \binom{n}{1}\binom{n}{1}\right]x^2(1-x)^{2n-2} \qquad (11)$$

$$p_3(x) = 2\left[\binom{n}{3} + \binom{n}{2}\binom{n}{1}\right]x^3(1-x)^{2n-3}$$

etc.

In principle we are free to choose the chain length 2n as long as the sum rule $\Sigma_m p_m(x)=1$ and the condition $2n \geq m$ are fulfilled. Fig. 10 displays the probabilities $p_m(x)$ for a chain of length 2n=24. The mean values $\langle p_m(x) \rangle$ of the probability distribution indicated by arrows scale linearly with the $Mn^{2+}$ concentration x as expected. It is tempting to assume that the linewidth of the dimer excitations also scales with $\langle p_m(x) \rangle$. Indeed, for small concentrations $x \leq 0.14$ the linewidth increases linearly with x as shown in Fig. 4. For larger linewidths, however, this linear relationship no longer holds. We find empirically that the linewidth follows a $\sigma^2$ law where $\sigma$ is the variance of the probability distributions indicated in Fig. 10.

TABLE I. Singlet-triplet splittings $\Delta_d$ and $\Delta_t$ for dimers and linear tetramers, respectively, in which the spins $\mathbf{s_i}$ are antiferromagnetically coupled by a nearest-neighbor exchange parameter J<0 (with vanishing biquadratic exchange parameter K=0). $\delta_{d,single-ion}$, $\delta_{t,single-ion}$ and $\delta_{d,two-ion}$, $\delta_{t,two-ion}$ denote the splittings of the first-excited triplet state |1> due to the anisotropy parameters D and $J^z$, respectively. The $\delta$ values listed in the table are calculated for very small anisotropy parameters |D|=|$J^z$|=0.01|J|, but they can be extrapolated up to |D|, |$J^z$| < 0.05|J| with a precision of at least 1%.

| $s_i$ | $\Delta_d$ | $\delta_{d,single-ion}$ | $\delta_{d,two-ion}$ | $\Delta_t$ | $\delta_{t,single-ion}$ | $\delta_{t,two-ion}$ |
|---|---|---|---|---|---|---|
| 1/2 | -2J | - | 1.00|$J^z$/J| | -1.318J | - | 1.56|$J^z$/J| |
| 1 | -2J | 1.00|D/J| | 2.00|$J^z$/J| | -1.018J | 0.95|D/J| | 3.52|$J^z$/J| |
| 3/2 | -2J | 2.40|D/J| | 3.40|$J^z$/J| | -0.952J | 2.76|D/J| | 6.77|$J^z$/J| |
| 2 | -2J | 4.20|D/J| | 5.20|$J^z$/J| | -0.946J | 5.44|D/J| | 11.3|$J^z$/J| |
| 5/2 | -2J | 6.39|D/J| | 7.39|$J^z$/J| | -0.950J | 8.90|D/J| | 17.0|$J^z$/J| |

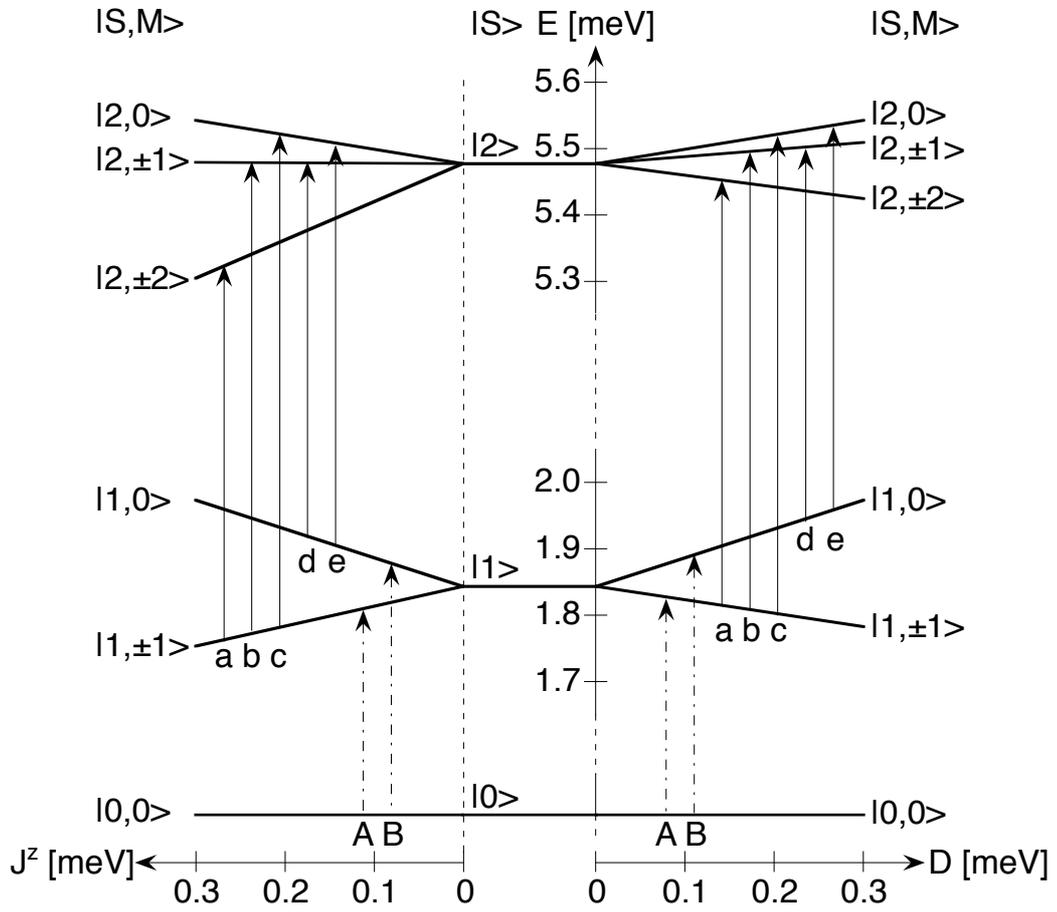

FIG. 1. Energy level splittings of dimers with $s_i=5/2$. The chosen energy scale corresponds to the $Mn^{2+}$ dimer splittings observed for $CsMn_xMg_{1-x}Br_3$ in the present work. The arrows mark the transitions allowed by the selection rules.

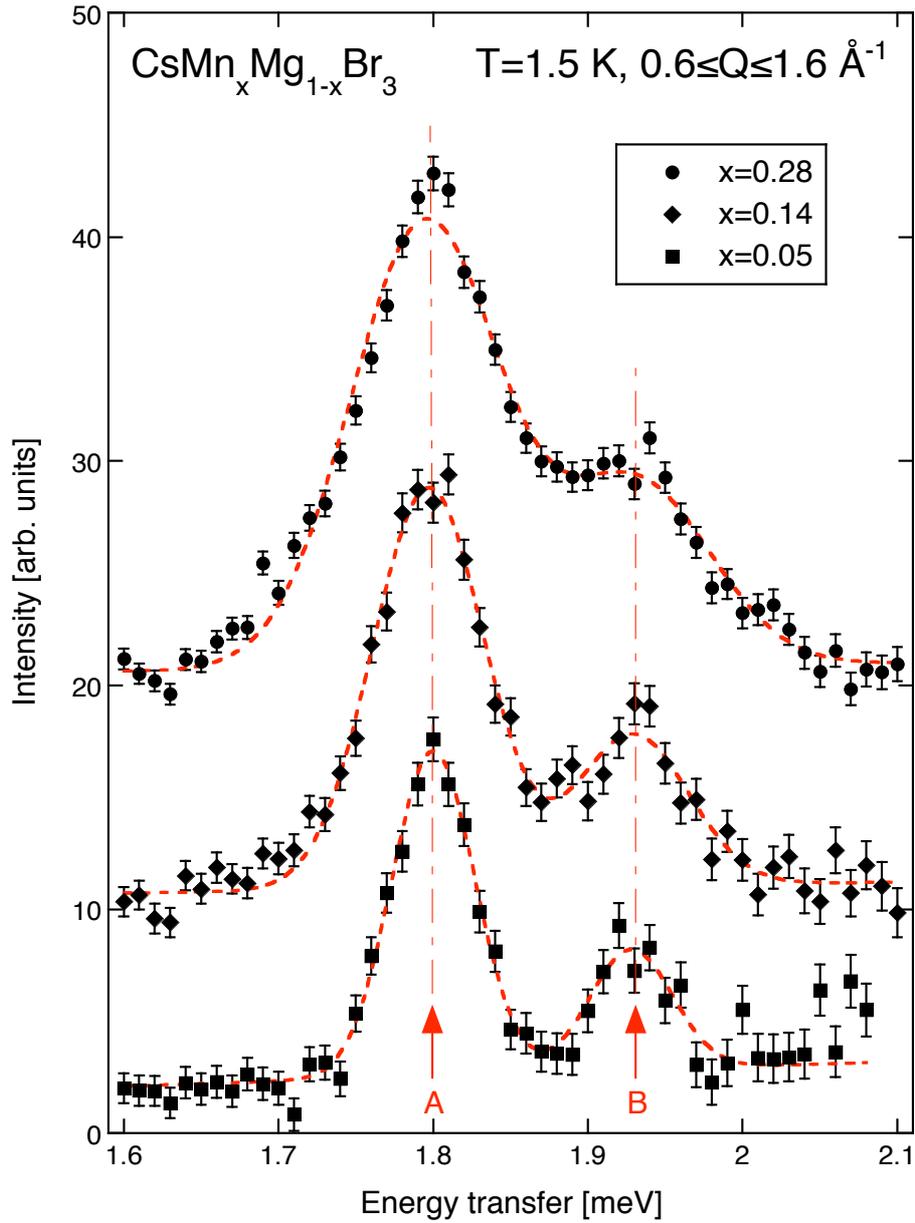

FIG. 2. (Color online) Energy spectra of neutrons scattered from $CsMn_xMg_{1-x}Br_3$ at T=1.5 K. The incoming neutron energy was 2.91 meV. For clarity, the data for x=0.14 and 0.28 are enhanced by 10 and 20 intensity units, respectively. The dashed lines refer to Gaussian peak fits with equal linewidths for both transitions. The arrows mark the transitions according to Fig. 1.

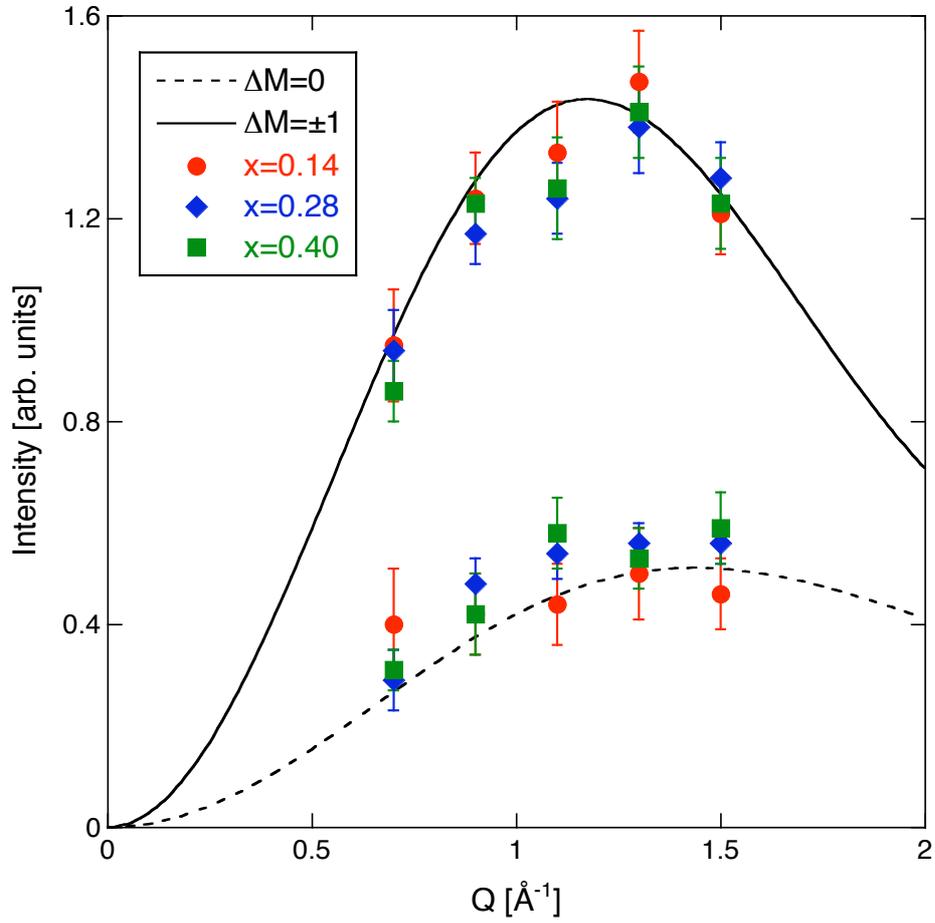

FIG. 3. (Color online) Q dependence of the neutron cross section for the $\Delta M=0$ and $\Delta M=\pm 1$ dimer transitions of $CsMn_xMg_{1-x}Br_3$. The symbols denote the intensities observed at T=1.5 K. The dashed and full lines correspond to the cross sections (6a) and (6b), respectively.

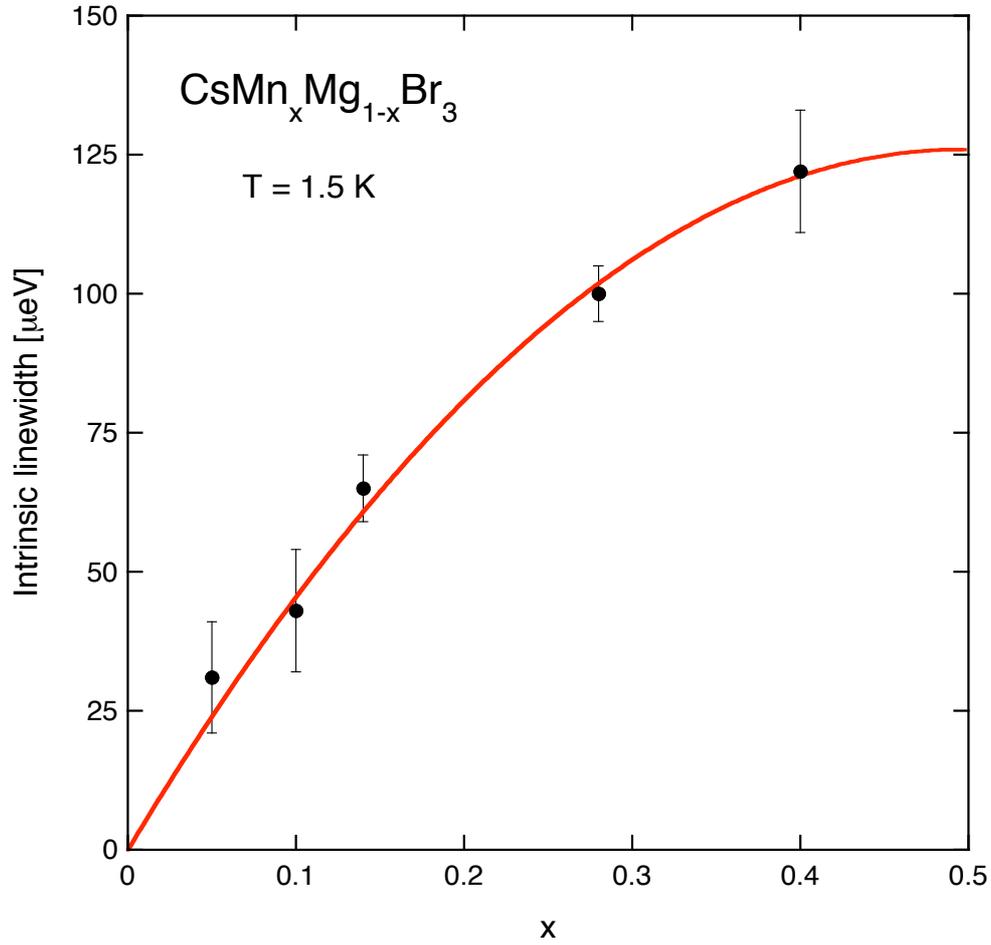

FIG. 4. (Color online) Intrinsic linewidths of the dimer transitions $|0,0\rangle \rightarrow |1,\pm 1\rangle$ and $|0,0\rangle \rightarrow |1,0\rangle$ observed for $CsMn_xMg_{1-x}Br_3$. The line corresponds to a $\sigma^2$ law as explained in the Appendix.

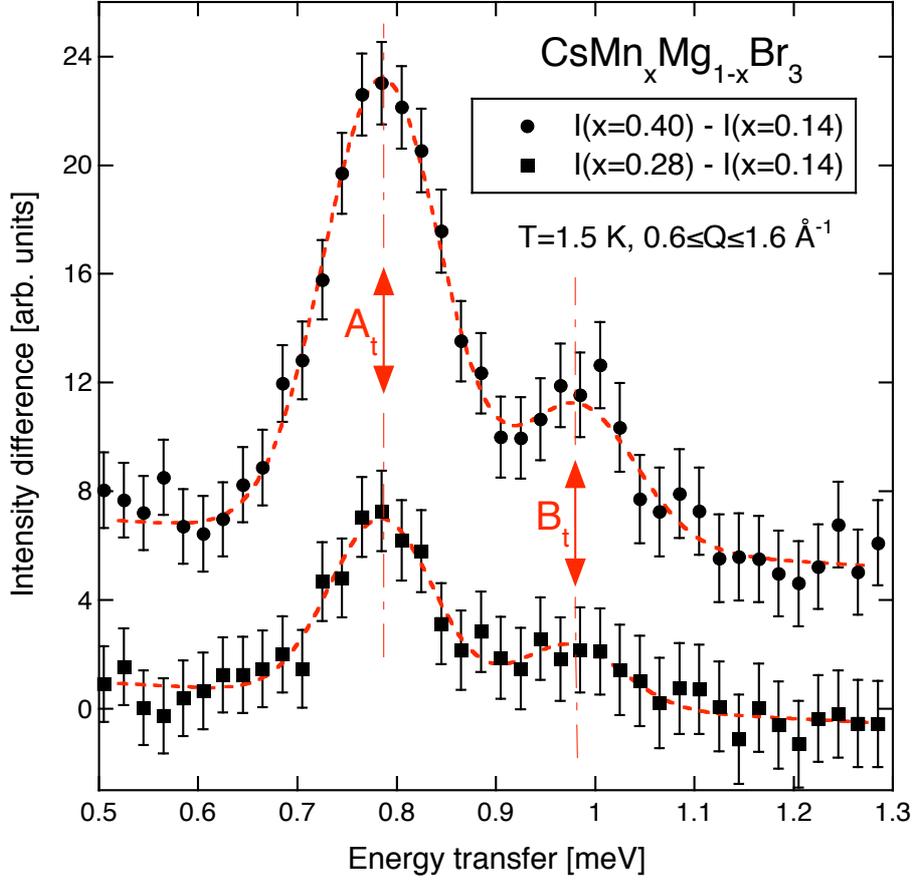

FIG. 5. (Color online) Energy spectra of neutrons scattered from $CsMn_xMg_{1-x}Br_3$ for x=0.28 and 0.40 at T=1.5 K, with subtraction of the x=0.14 data. The incoming neutron energy was 2.91 meV. For clarity, the data for x=0.40 are enhanced by 6 intensity units. The arrows marked by $A_t$ and $B_t$ denote the tetramer transitions $|0,0\rangle \rightarrow |1,\pm1\rangle$ and $|0,0\rangle \rightarrow |1,0\rangle$, respectively. The dashed lines refer to Gaussian peak fits with equal linewidths for both transitions.

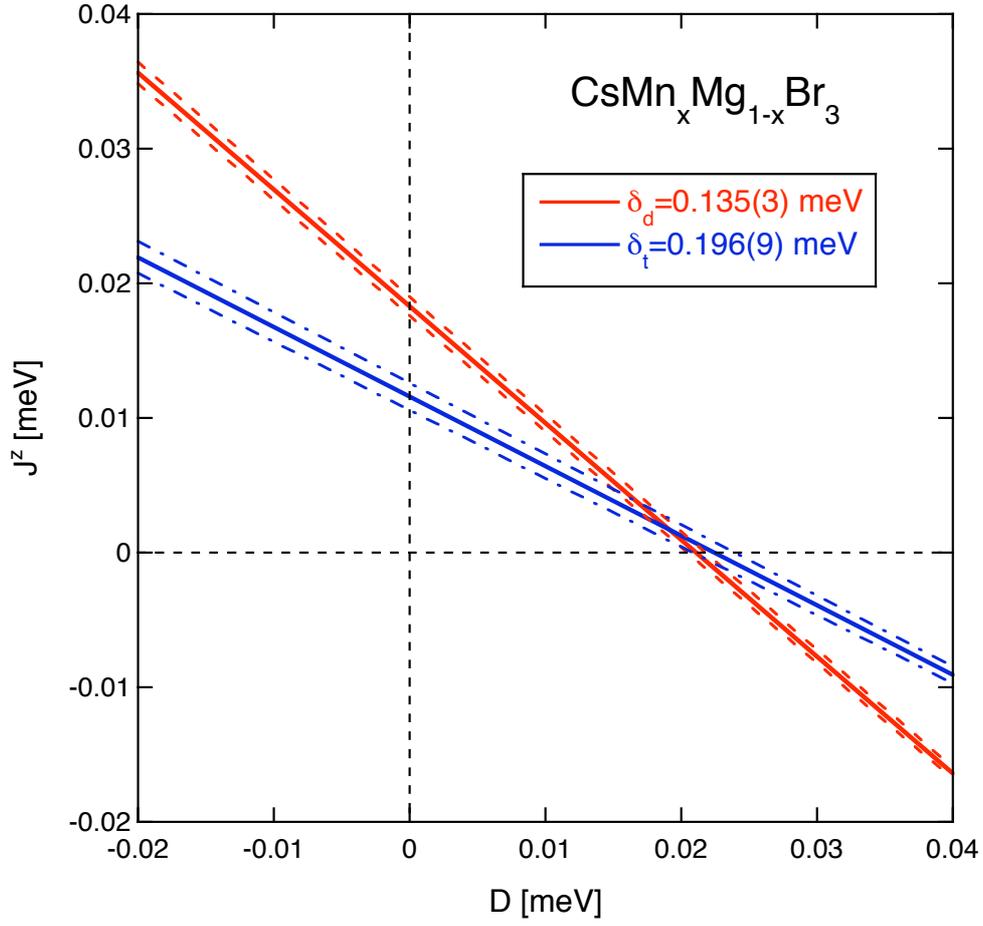

FIG. 6. (Color online) Plot of the anisotropy parameters D and $J^z$ compatible with the observed triplet splittings $\delta_d$ and $\delta_t$. The dashed and dashed-dotted lines indicate the experimental uncertainties associated with $\delta_d$ and $\delta_t$, respectively.

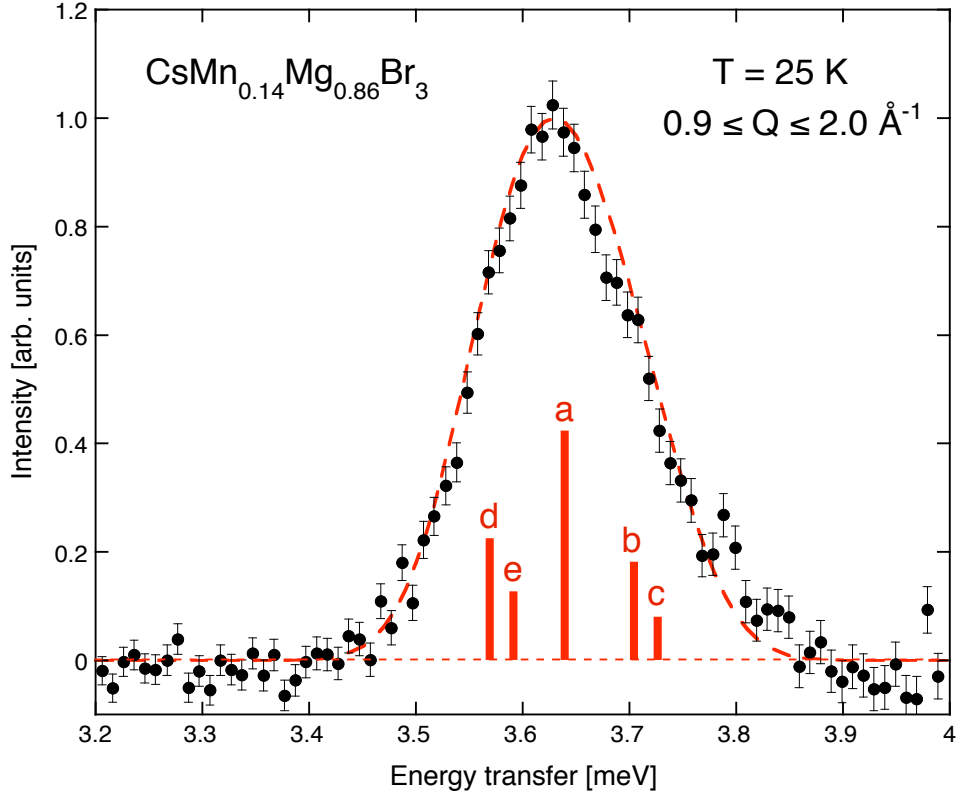

FIG. 7. (Color online) Energy spectrum of neutrons scattered from $CsMn_{0.14}Mg_{0.86}Br_3$ at T=25 K. The incoming neutron energy was 5.11 meV. The dashed line corresponds to a superposition of five Gaussians corresponding to the five allowed triplet-quintet transitions whose energies and intensities (calculated from the model parameters) are indicated as vertical bars and numbered according to Fig. 1. The linewidth of the individual Gaussians was set at the instrumental resolution of 120 μeV.

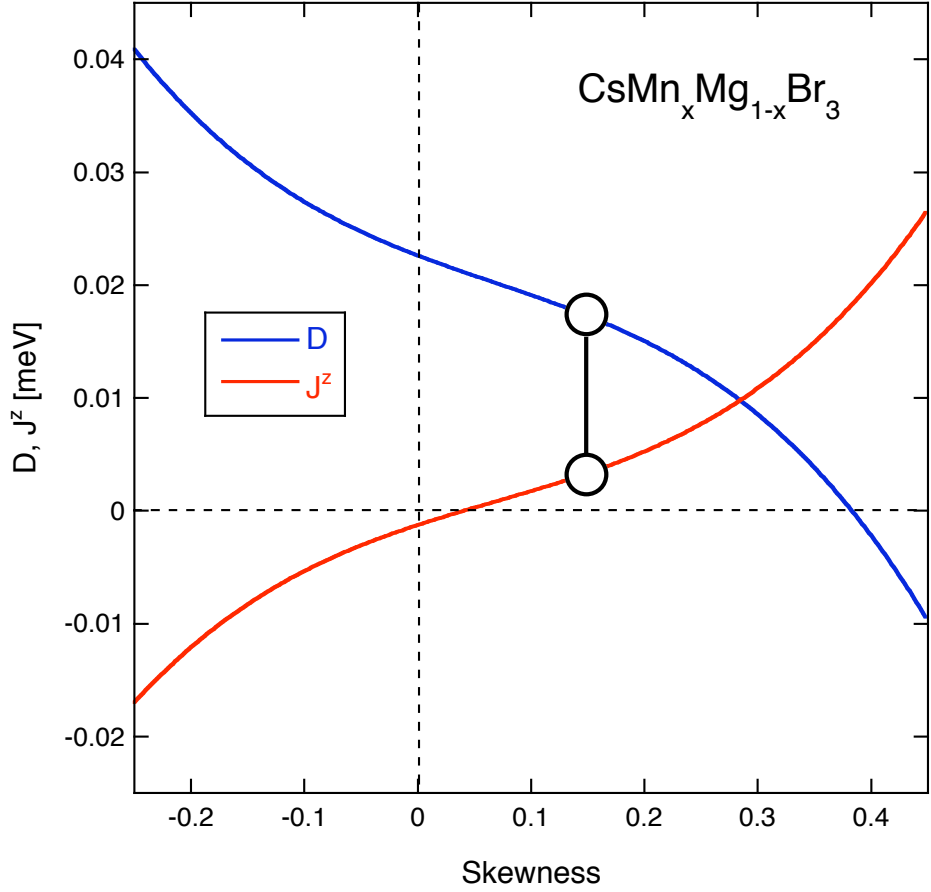

FIG. 8. (Color online) Skewness of the dimer excitation $|1\rangle \to |2\rangle$ of $CsMn_xMg_{1-x}Br_3$ calculated for $(D,J^z)$ pairs compatible with the observed splitting $\delta_d=0.135$ meV of the dimer triplet state according to Fig. 6. The calculation is based on a convolution of the five allowed transitions a-e indicated in Fig. 1 with an instrumental resolution of 120 μeV. The dumbbell marks the $(D,J^z)$ pair derived from the energy spectrum of Fig. 7 which gave a skewness of s=0.155(19).

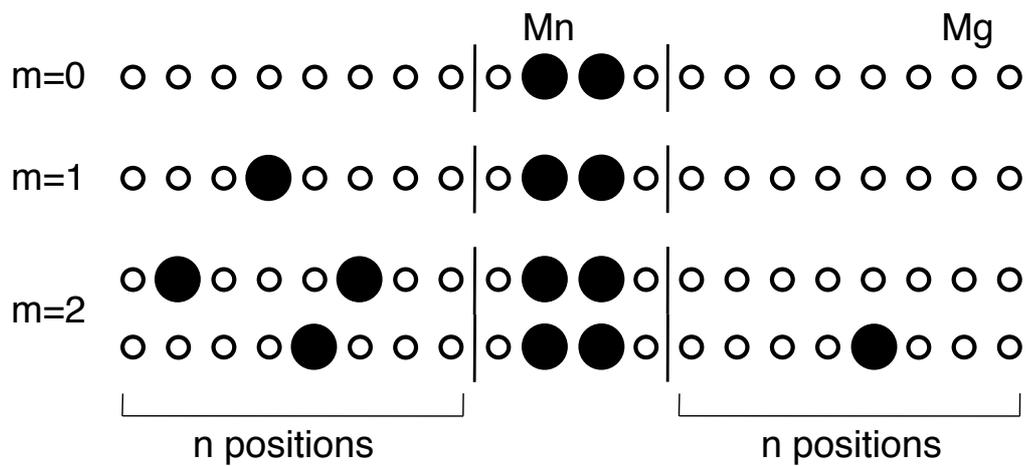

FIG. 9. Sketch of the atomic configurations along a mixed $Mn_xMg_{1-x}$ chain. $2n$ denotes the chain length outside the central $Mn^{2+}$ pair, and $m$ is the number of peripheric $Mn^{2+}$ ions in the chain.

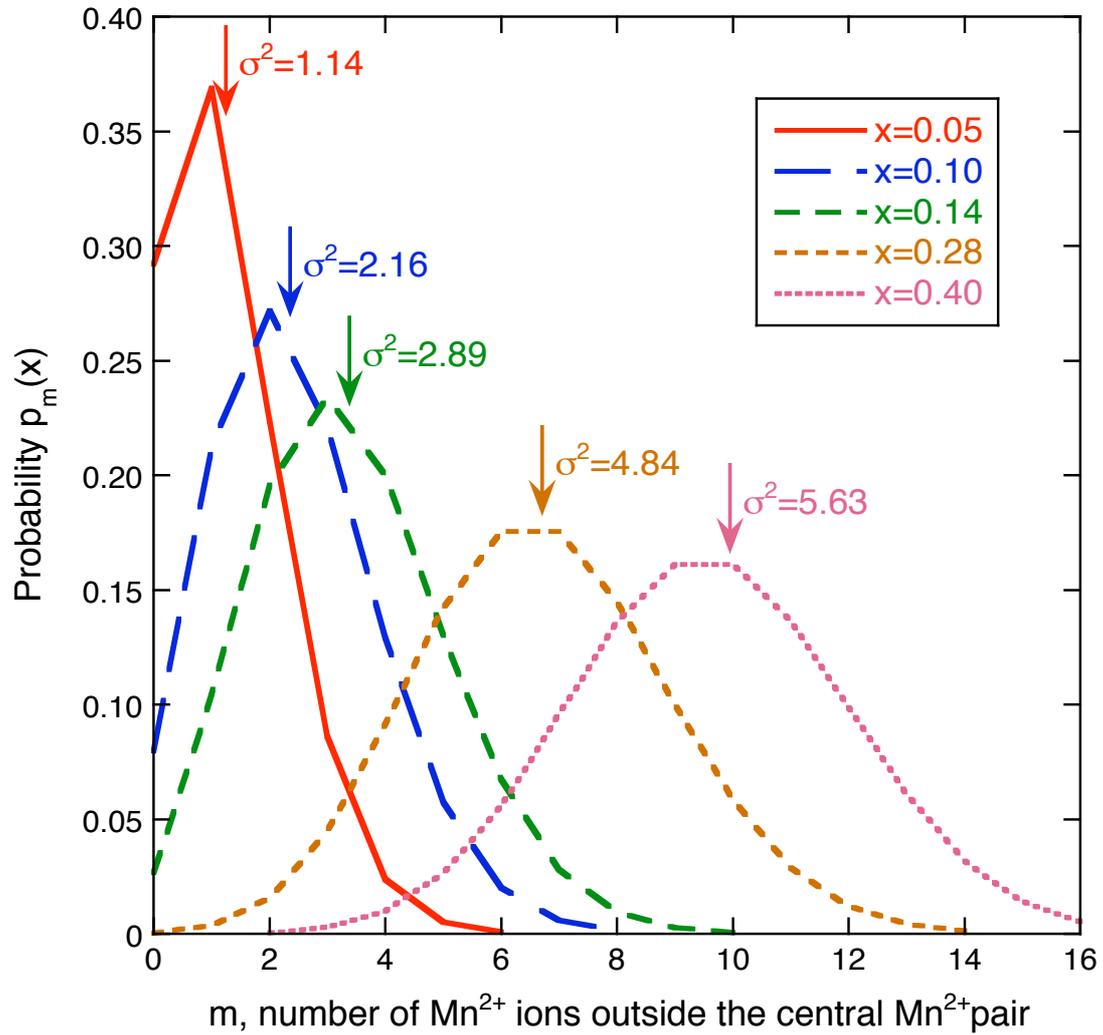

FIG. 10. (Color online) Distribution of the probabilities $p_m(x)$ for a mixed $Mn_xMg_{1-x}$ chain of length $2n=24$. The arrows mark the mean values $\langle p_m(x)\rangle$, and the square of the variance $\sigma$ is also indicated.